\documentclass[journal]{IEEEtran}
\usepackage{multirow}
\usepackage{color}
\usepackage{indentfirst}
\usepackage{mathrsfs}

\usepackage{amsmath}
\ifCLASSINFOpdf
  \usepackage[pdftex]{graphicx}
  \usepackage[justification=centering]{caption}
  \graphicspath{{./figures/}{./biographies/}}
  \DeclareGraphicsExtensions{.pdf,.jpeg,.png,.jpg}
\else
\fi

\ifCLASSOPTIONcompsoc
  \usepackage[caption=false,font=normalsize,labelfont=sf,textfont=sf]{subfig}
\else
  \usepackage[caption=false,font=footnotesize]{subfig}
\fi
\usepackage{algorithm}
\usepackage{algorithmic}

\usepackage{cite}

\begin{document}
%
\title{Low-light Image Restoration with Short- and Long-exposure Raw Pairs}
%
%
%

\author{Meng Chang,
        Huajun Feng,
        Zhihai Xu,
        Qi Li
\thanks{M Chang, H Feng, Z Xu and Q Li are with the State Key Laboratory of Modern Optical Instrumentation, Zhejiang University, Hangzhou, Zhejiang, 310027 China e-mail:liqi@zju.edu.cn.}
}

\markboth{Journal of \LaTeX\ Class Files,~Vol.~14, No.~8, August~2015}%
{Shell \MakeLowercase{\textit{et al.}}: Bare Demo of IEEEtran.cls for IEEE Journals}
%



\maketitle

\begin{abstract}
Low-light imaging with handheld mobile devices is a challenging issue. Limited by the existing models and training data, most existing methods cannot be effectively applied in real scenarios. In this paper, we propose a new low-light image restoration method by using the complementary information of short- and long-exposure images. We first propose a novel data generation method to synthesize realistic short- and long-exposure raw images by simulating the imaging pipeline in low-light environment. Then, we design a new long-short-exposure fusion network (LSFNet) to deal with the problems of low-light image fusion, including high noise, motion blur, color distortion and misalignment. The proposed LSFNet takes pairs of short- and long-exposure raw images as input, and outputs a clear RGB image. Using our data generation method and the proposed LSFNet, we can recover the details and color of the original scene, and improve the low-light image quality effectively. Experiments demonstrate that our method can outperform the state-of-the-art methods.
\end{abstract}

\begin{IEEEkeywords}
Low-light imaging, denoising, deblurring, image fusion, image restoration.
\end{IEEEkeywords}

%
\IEEEpeerreviewmaketitle

\section{Introduction}

\IEEEPARstart{I}{maging} in a low-light environment is always a challenging subject, especially on mobile devices. Limited by the quality of optical imaging devices on these devices, images captured in low light suffer from high level noise, low visualization and color distortion when the exposure time is short. One way to reduce noise and obtain accurate color is extending the exposure time with sensor sensitivity reduction. However, due to the camera motion of handheld devices or object motion in the scenes, the resulting long-exposure images suffer from motion blur. In addition, many highlighted areas may have a large area of overexposure, causing the dynamic range of the image to be cut off. Therefore, to obtain high-quality images in dark scenes, image postprocessing is often required.

Most existing methods focus mainly on one aspect of image denoising~\cite{dabov2007image, dabov2007color, zhang2017beyond, zhang2018ffdnet, guo2019toward, anwar2019ridnet, chen2018learning, jin2020a, 9136787} or deblurring~\cite{xu2010two,pan2016blind,pan2016robust,nah2017deep ,tao2018scale,ren2019adjusted,Zhang_2019_CVPR}. Noise removal methods can obtain sharp images, avoiding overexposure. Hence, many low-light enhancement methods capture short-exposure images, and then they increase the brightness and reduce the noise to produce normal-exposure images. This method can allow for a higher dynamic range and avoidance of motion blur. However, the true colors and missing details of the scene are hard to recover. On the other hand, because the motion of the scene is complex, the blur in the image is not uniform. Motion blur is more difficult to deal with than noise especially when there are outliers in low-light scenes. The presence of an overexposed area near a light source may mislead blur removal tasks. Deblurring methods cannot deal with dark scenes well and always result in blurred residuals or other artifacts. Another approach is to combine the advantages of both long/short-exposed images into one high-quality image. Many previous methods have been studied to achieve this goal, and some progress has been made~\cite{yuan2007image, rengarajan2019photosequencing, mustaniemi2018lsd, zhang2019deep}. However, limited by their algorithms or the data generation method, these methods cannot be effectively applied to images taken by real handheld mobile devices.

The existing methods of synthesizing long- or short-exposure images are often straightforward. For short-exposure images, they add Gaussian noise directly to the standard RGB (sRGB) domain to obtain noise images. Additionally, for long-exposure images, uniform or nonuniform blur kernels are used to generate blurred images in the sRGB domain or after inverse gamma correction. These methods do not take the formation of noise or blur into account in the imaging pipeline. The images generated by these methods are only approximations of the real images and ignore many factors in the imaging process. Recently, many methods have been proposed to generate noise images by taking real dark noise images~\cite{abdelhamed2018high, chen2018learning} or by simulating the noise factors in the imaging process~\cite{abdelhamed2019noise, wang2019enhancing, wei2020physics}, and great progress has been made. Limited by the information of a single noisy image, however, the performance remains unsatisfactory. In addition, there is no method to analyze the generation of long-exposure blurred images from the perspective of a real imaging pipeline.

In this paper, we simulate a real low-light imaging pipeline and propose a novel method to synthesize long/short-exposure images. The camera sensor actually receives raw images, which are then processed by an ISP pipeline to produce visual sRGB images. Therefore, our data generation method first synthesizes raw images with long and short exposures. Then, to avoid the damage to image information caused by ISP operation, we directly use the long- and short-exposure raw image to synthesize a high-quality image.

There are three main challenges in image fusion: misalignment, ghosting, and information fusion.  First, since the images are acquired consecutively, the time delay causes displacement between them. Second, areas that are misaligned or inconsistent between images result in ghost artifacts after fusion. Third, the fusion method needs to extract useful information from different images, and then improve the visual effects and fidelity. To achieve this goal and avoid the aforementioned limitations, we propose an long-short-exposure fusion network (LSFNet) to fuse long/short-exposure images. LSFNet consists of three designed modules to address each of the issues described above. To improve the details, a multiscale structure is used to reconstruct the image from coarse to fine. The experimental results demonstrate that our method is effective at improving low-light image quality.

 In conclusion, the main contributions of our work are as follows:
\begin{itemize}
    \item Based on a highly accurate imaging model in a low-light environment, we propose a data generation approach that synthesizes long/short-exposure raw images. With the contribution of our dataset, we are able to better deal with real low-light images obtained with mobile devices.

    \item We propose a novel long-short exposure fusion network architecture, which takes the long/short-exposed raw images as input. The proposed network can handle misalignment, ghost artifacts and information fusion issues, and then outputs a high-quality RGB image.

    \item We compare our method with various low-light enhancement methods on synthetic and real test datasets. The results demonstrate that our model achieves state-of-the-art performance on both synthetic and real images with vivid visual effects.
\end{itemize}

\section{Related Works}
The previous work focused mainly on two aspects, deblurring and denoising, to improve image quality. Deblurring and denoising are two classic ill-posed problems and have been studied extensively.

Traditional single-image denoising attempts to model the distribution of natural image or noise, and using this prior information to recover clear images with optimization algorithms. The common priors include sparsity~\cite{aharon2006k, xu2018trilateral}, non-local self-similarity~\cite{dabov2007image,dabov2007color, gu2014weighted} and external statistical prior~\cite{roth2009fields,zoran2011learning,xu2018external}. Recently, learning-based methods have been proposed to remove noise from end to end ~\cite{zhang2017beyond, zhang2018ffdnet,jin2020a} and improve the performance on more realistic noise~\cite{guo2019toward,anwar2019ridnet,9136787}. Moreover, great progress has been made in the acquisition of noise data. Many recent works have begun to analyze the source of noise from the perspective of image acquisition~\cite{abdelhamed2019noise, wang2019enhancing, wei2020physics}, and remove noise from raw images~\cite{chen2018learning, brooks2019unprocessing}. Because raw images retain more information than RGB images, with simpler noise distributions, image recovery tasks tend to yield better results on raw images.

Similar to image denoising methods, image deblurring methods can be divided into model-based methods and learning-based methods. Model-based methods often assume that the image has uniform blur, and introduce prior information to suppress the noise and ringing artifacts~\cite{xu2013unnatural, pan2016blind, pan2016robust, ren2019adjusted}. The learning-based methods use convolutional networks to process blurred images and output sharp images directly. Learning-based approaches can deal with nonuniform blur since they do not require explicit estimation of the blur kernel~\cite{nah2017deep,tao2018scale,zhang2019deep}. However, single-image deblurring methods are difficult to have good robustness due to the diversity of motion blur. On the other hand, it is difficult to capture blur-sharp image pairs in real scenes, so the existing blur datasets are synthesized, including convoluting uniform/nonuniform blur kernels~\cite{kohler2012recording, lai2016comparative} or averaging consecutive short-exposure frames from high-frame-rate videos~\cite{nah2017deep}. These methods all add blur to the sRGB domain without considering the formation of blur from the low-light imaging pipeline, which limits their recovery performance.

In addition to single-image restoration, burst image denoising~\cite{hasinoff2016burst, godard2018deep, mildenhall2018burst} or deblurring methods~\cite{wieschollek2017learning, aittala2018burst} have also made much progress. Burst image methods consecutively capture multiple images with the same exposure and fuse them together to improve the image quality. However, to achieve the ideal signal-to-noise ratio, it is often necessary to acquire many frames of images, which increases the difficulty of image registration. On the other hand, limited by the constant exposure, complementary information is missing for accurate recovery of high-dynamic-range scenes.

Noisy-blurred image pairs have been used in image deblurring tasks~\cite{yuan2007image, rengarajan2019photosequencing,zhang2019deep}. Using the texture of the noisy image as a constraint, these pairs can effectively removes severe blur and suppress ringing artifacts.
Short-long-exposure fusion methods~\cite{Mertens2007Exposure, kalantari2017deep} were previously used to improve the dynamic range of images, without considering noise and blur in the imaging process. Recently, LSD$_2$~\cite{mustaniemi2018lsd} was used to combines long- and short-exposure images for joint noise and blur removal, to improve the quality of low-light images. This approach uses only a channel-augmented UNET architecture and takes the RGB images after ISP postprocessing as input, which cannot obtain a satisfactory result in real images captured by mobile devices.

\begin{figure*}[t]
\begin{center}
   \includegraphics[width=0.85\linewidth]{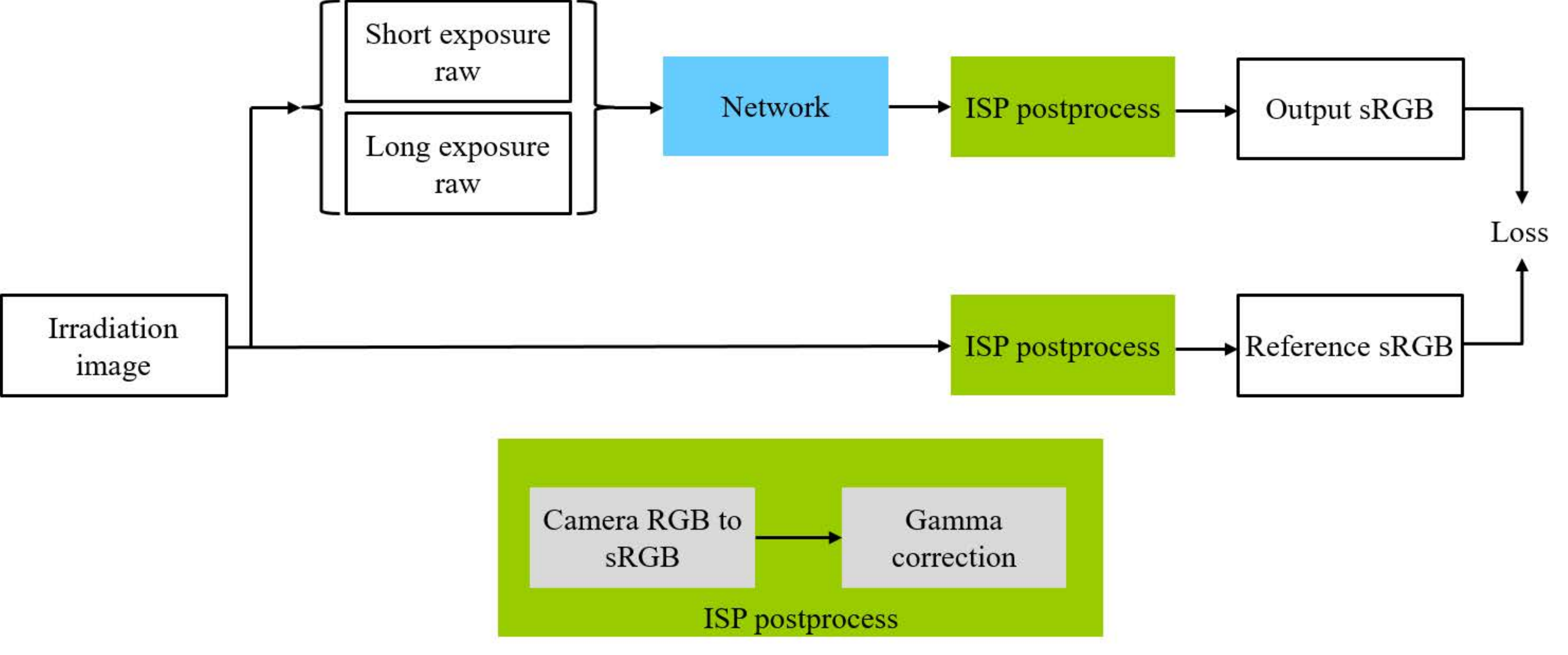}
\end{center}
   \caption{The overview of our method.}
\label{fig:overview}
\end{figure*}

\begin{figure}[t]
\begin{center}
   \includegraphics[width=1\linewidth]{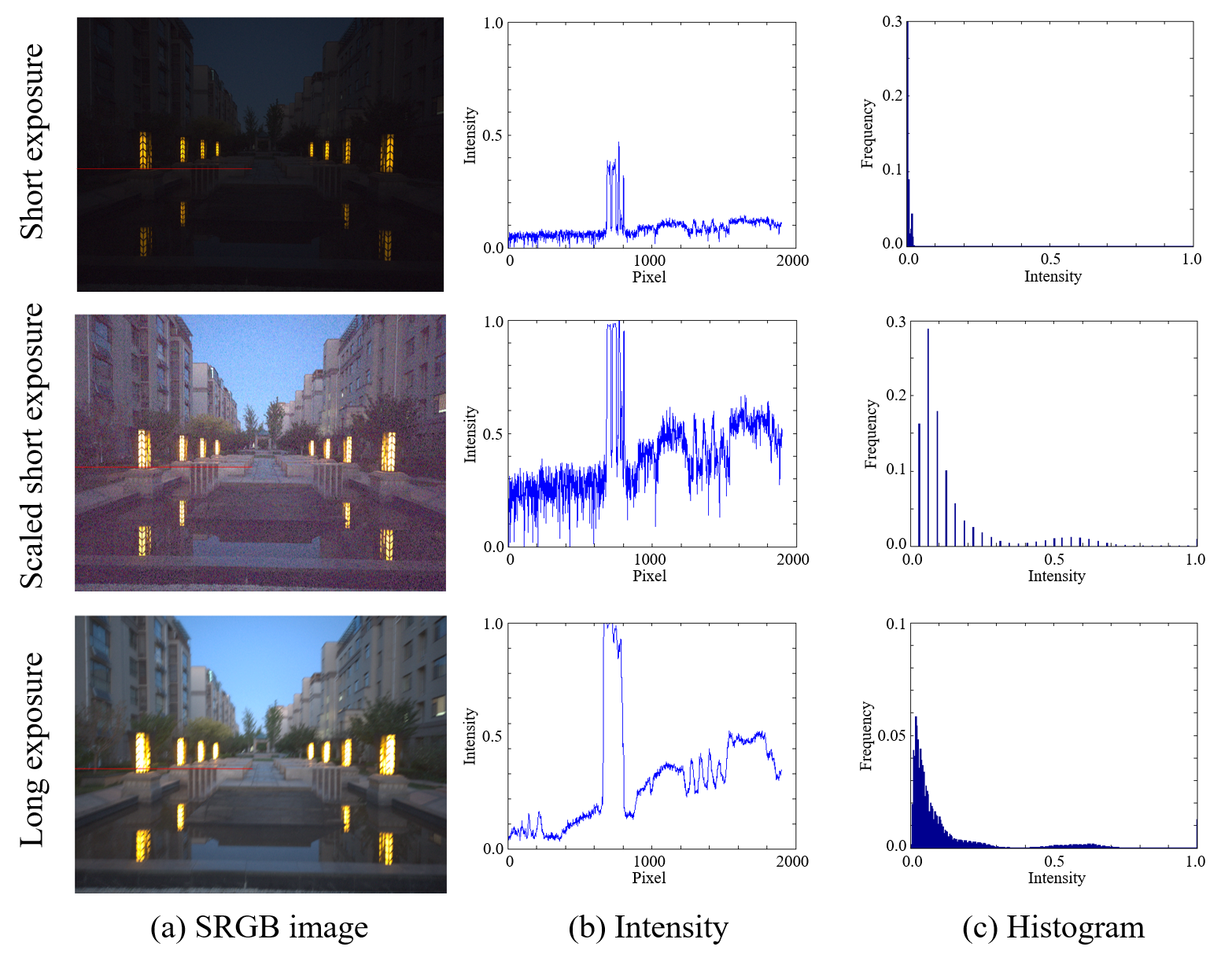}
\end{center}
   \caption{\textbf{Intensity maps and histograms for short- and long-exposure images.} (a) sRGB images for different exposures. (b) Intensity values of pixels in the red line. (c) Histograms of the raw images.}
\label{fig:hist}
\end{figure}

\section{Background}
As mentioned above, imaging in a low-light environment is a challenging issue.  Different exposure strategies always have their own drawbacks. Short exposure times with high ISO values lead to high noise levels and color distortion. A long exposure time with low ISO results in motion blur. Both exposure strategies may have a cutoff dynamic range limited by the camera sensor. Compounding the problem, furthermore, is the fact that these flaws are often not independent of each other.

First, noise is inevitable in the process of imaging. Even if we can reduce the noise by increasing the exposure time and lowering the ISO value, the noise cannot be completely eliminated, especially in low-light environments. This can result in both noise and motion blur existing in the images. Denoising and deblurring tasks are often mutually constrained, which may amplify noise when removing blur. Second, motion blur causes the overexposure outliers to form sharp edges, which may mislead the deblurring task~\cite{cho2011handling}. Moreover, the large areas of overexposure are not conducive to the removal of blur and noise.

The interaction between a high noise level and exposure cutoff is also one of the reasons for color distortion in short-exposure images. The red and blue values are always smaller than the green values in captured raw images. Due to high noise level, more pixels in the red and blue channels fall below the dark current value and are cut off, resulting in the red and blue channels having higher brightness levels after white balancing.

Noise and blur also limit the expansion of the dynamic range. The common way to extend the dynamic range is multi-exposure fusion. This method extracts the information of normal-exposure regions in different-exposure images and then generates a high-dynamic-range image. However, noise and other artifacts are fused as the textures of the images, severely degrading the fused image quality. Another way to obtain high-dynamic-range images is to capture short-exposure images while avoiding overexposure, and then use tone mapping to increase the brightness. However, the increase in brightness amplifies noise at the same time, as shown in Fig.~\ref{fig:hist}. In addition, to obtain the high dynamic range of real scenes, most dark regions are often compressed. Limited by the quantification scope of the sensor, many details in dark regions are quantitatively compressed into a small bit range.
Fig.~\ref{fig:hist} (c) shows the histograms of raw images for different exposures. Compared to the long-exposure image, the intensity values of the scaled short-exposure image are quantified in a few discrete levels. Moreover, color correction is also a great challenge with short-exposure images.

Therefore, simple denoising or deblurring operations alone cannot effectively improve the quality of low-light images and sometimes even make the results worse. We need to take all factors into consideration and use the complementary information of short and long exposures to improve the image quality.

\section{Approach}

Short/long-exposure images generation is the key task in learning-based methods. Most existing methods generate blur images only by convoluting uniform/nonuniform blur kernels and adding homoscedastic Gaussian noise in the sRGB domain. However, this is only an approximation of the real imaging process. In fact, the blurred and noisy images captured by the camera are processed by the ISP system. Demosaicing operations make pixels independent, and some nonlinear operations, such as gamma correction or tone mapping, also affect the imaging results. Therefore, we generate long- and short-exposure images in the raw domain, which is unaffected by ISP operations and scaled linearly to the intensity received by the sensor. In practice, short- and long-exposure images can be acquired by the burst mode of mobile devices. Short-exposure inputs have shorter exposure times than long-exposure inputs. Similar to \cite{mustaniemi2018lsd}, we choose an exposure ratio of 30, which means that the short exposure time is 1/30 of the long exposure time. We scale the short-exposure image to match the brightness when being input into the network.

The proposed LSFNet takes the long- and short-exposure raw images as input, and outputs a camera RGB image, which is sharp, noise free and color corrected. Then, the ISP postprocessing transforms the output to a vivid sRGB image. An overview of our method is shown in Fig.~\ref{fig:overview}. In the following section, we will give the details on our data generation method and the architecture of LSFNet.

\begin{figure*}[t]
\begin{center}
   \includegraphics[width=0.8\linewidth]{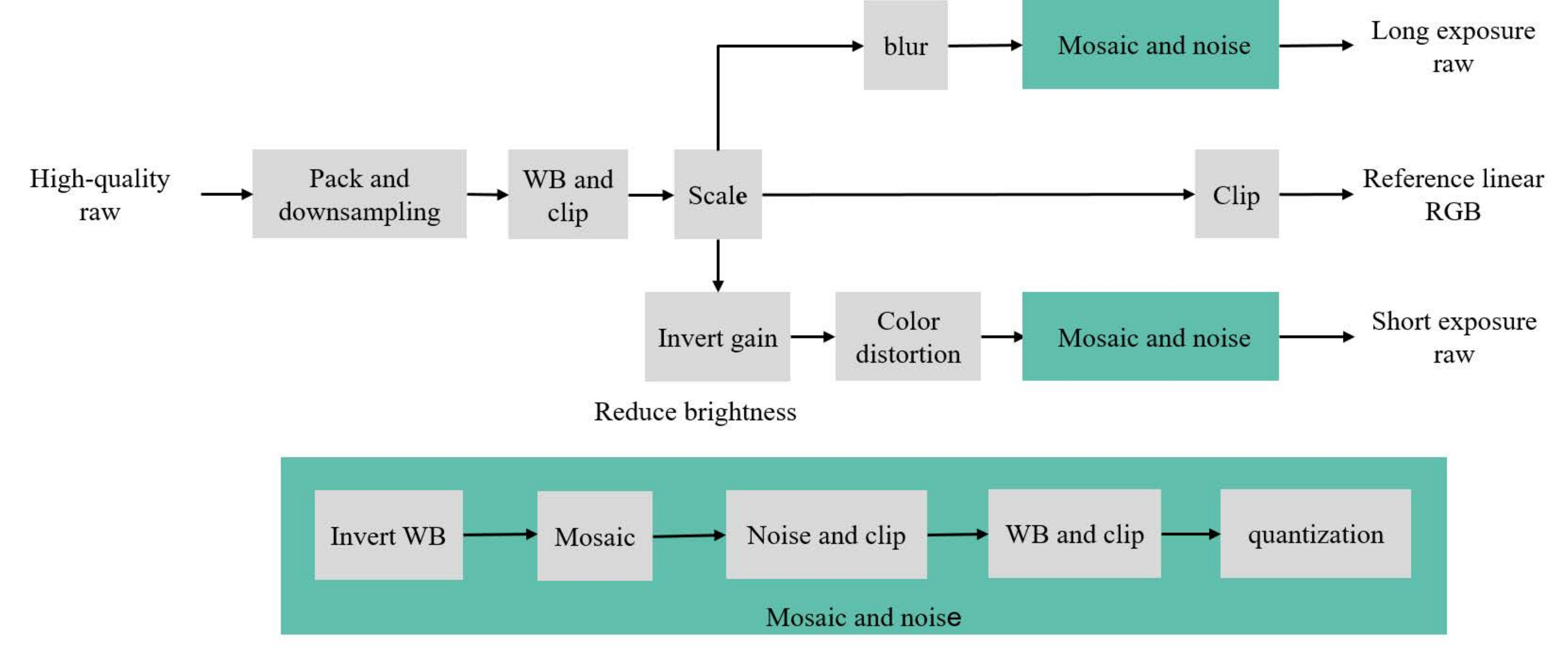}
\end{center}
   \caption{The overview of our data generation method.}
\label{fig:data_gen}
\end{figure*}

\begin{figure}[t]
\begin{center}
   \includegraphics[width=1\linewidth]{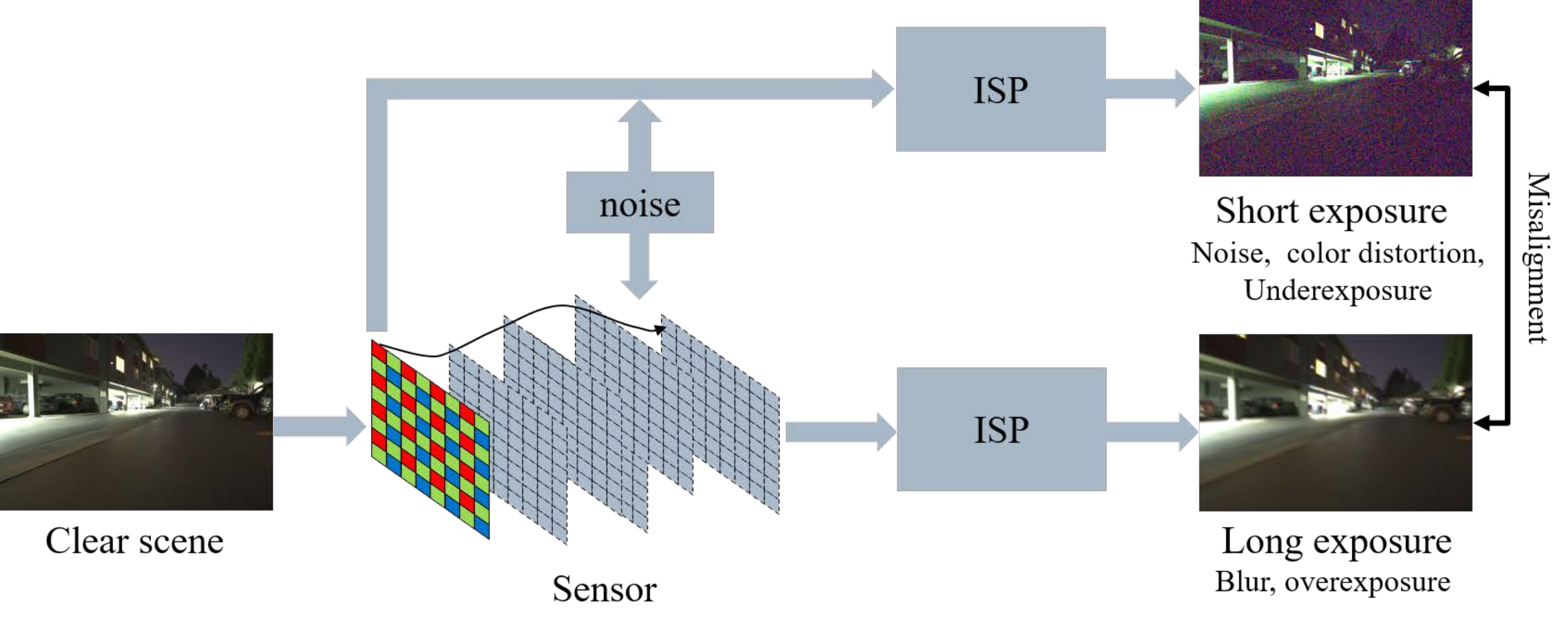}
\end{center}
   \caption{The imaging pipeline.}
\label{fig:imaging_pipe}
\end{figure}

\section{Data Generation}

The training data include pairs of short/long-exposure images with the corresponding high-quality clear images. The short-exposure images are sharp but noisy and color distorted. In addition, underexposed areas and quantization artifacts may also exist in dark regions. On the other hand, the long-exposure image is blurred, with areas of overexposure. Moreover, consecutive imaging with handheld devices will inevitably introduces inter-frame misalignment. We need to introduce these factors in turn when simulating the imaging process to synthesize data.

The imaging process is shown in Fig.~\ref{fig:imaging_pipe}. The camera sensor receives the irradiation from the scene, and obtains Bayer-pattern raw images through a color filter. When we make a long exposure, the camera shake overlays the irradiation of different scenic spots in the same pixel position of the sensor. Therefore, motion blur is generated in the irradiation domain. Then, the color filter samples the irradiation received by the sensor to form the blurred raw images. At the same time, noise is introduced due to the random fluctuation of photons and the imperfect nature of sensor devices. The blurred and noisy raw data pass through the ISP pipeline in the camera to obtain sRGB blurred images, constituting the final output of the camera. Therefore, in order to synthesize more realistic blur and noise data, we first need to obtain the irradiation domain images received by the camera sensor.

However, what we can obtain directly from the camera are only sampled Bayer-pattern raw images or the sRGB images processed by the ISP pipeline. To solve this problem, we first collect some high-quality raw images without noise and blur, and try to avoid large overexposed and underexposed areas. Then, we split the high-quality raw image with the Bayer pattern into three color channels and use the maximum entropy downsampling operation proposed by \cite{khashabi2014joint} to align each color channel, which leads to each pixel has three color values. In this way, we can synthesize the irradiation images $I_r$  received by the sensor. We subsequently add blur and noise to the irradiation images to obtain the low and short exposure raw data. The overview of our data generation method is shown in Fig.~\ref{fig:data_gen}.

\subsection{Overexposure outliers}
The real low-light scenes may include light sources that lead to some overexposed areas. Especially in the light source areas, even the short-exposure images are partially overexposed. Therefore, to simulate different overexposures, we allow a small number of light sources to exist in the irradiation images and exposure cutoffs to exist at the light source points. Then, we increase the whole brightness of the irradiation images by multiplying them by a scale factor $s$, which is uniformly sampled from the range [1.3, 3]. For the ground truth images, we can directly clip the scaled irradiation images $sI_r$ to the range of $[0,1]$ to obtain the long-exposure sharp images. For the long- and short- exposure inputs, we should clip to this range after adding motion blur and noise.

To simulate the light source regions, when producing short-exposure images, we keep the brightness of the cutoff regions unchanged and reduce the brightness of the other regions by dividing the exposure ratio $r$, which is set to 30, similar to the method in \cite{brooks2019unprocessing}. This allows areas of the light source that are cut off in the original irradiation images $I_r$ to remain cut off in the short-exposure images, while the scaled highlight areas in long-exposure images $sI_r$ are normally exposed.

\begin{figure}[t]
\begin{center}
   \includegraphics[width=1\linewidth]{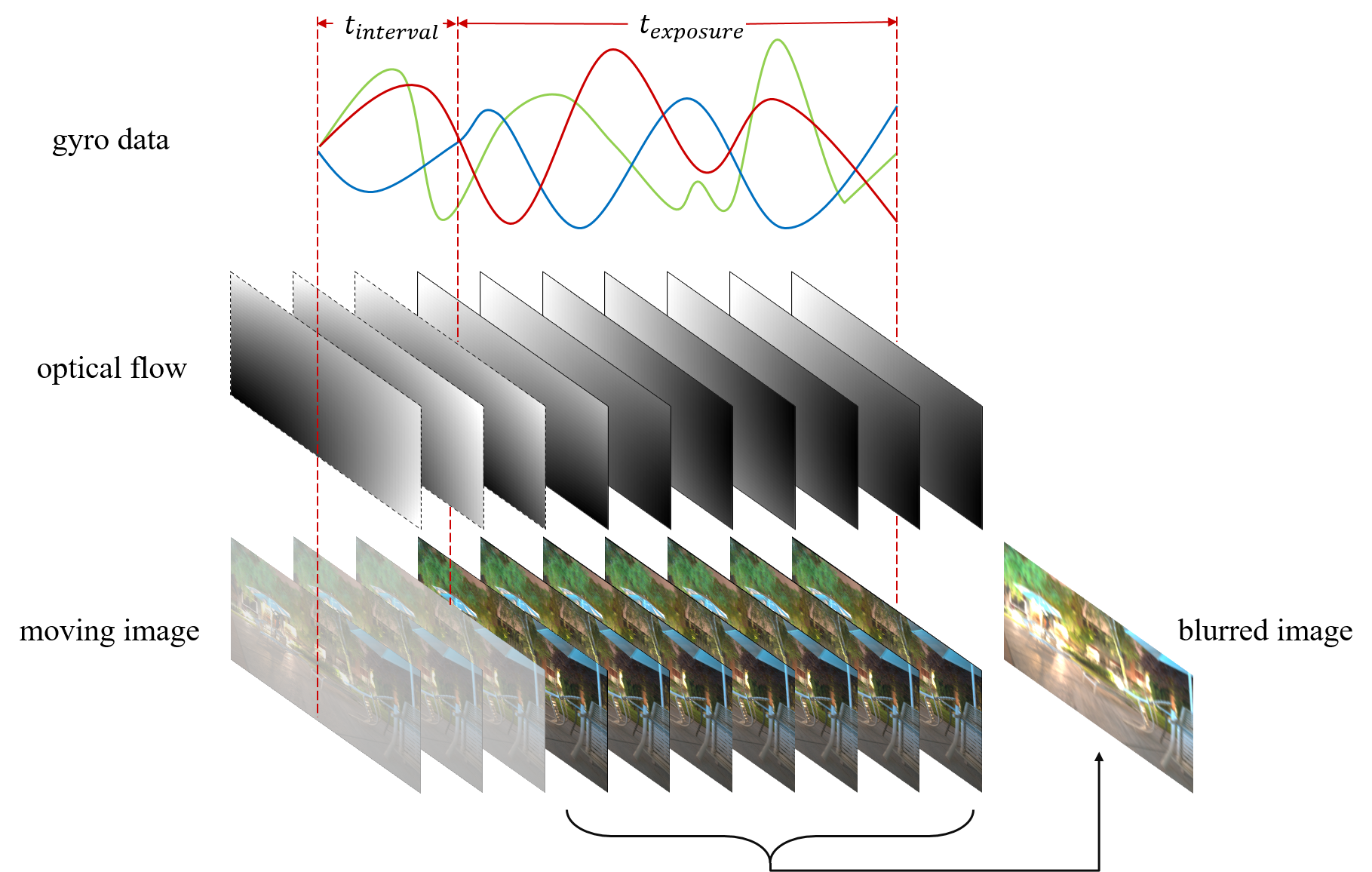}
\end{center}
   \caption{Long-exposure blurred image synthesis method.}
\label{fig:blur_gen}
\end{figure}

\subsection{Synthesis of long-exposure raw images}

Motion blur is the main cause of long-exposure image degradation. As mentioned above, motion blur should be added to the scaled irradiation images $sI_r$. We focus on blur caused by camera motion, which can be described from an imaging model~\cite{sindelar2013image, hee2014gyro}. To generate more realistic motion blur, we obtain camera motion information by recording the built-in gyroscope data. Then, an optical flow map is generated according to the imaging model, which indicates how each pixel is moving at each moment. We can generate the irradiation images captured by the sensor at each moment. When generating multiple warped frames, we ensure that the displacement between adjacent frames does not exceed one pixel, so that the resulting blurred images are continuously changing. By averaging the moving image series over an exposure time, we can obtain the long-exposed blur image.

We assume that short- and long-exposure images are shot consecutively, and the scene of the ground truth image should be consistent with the short exposure image. In generation, each pixel continuously moves from the position of the original irradiation image. In addition to the camera motion during long exposure, the motion in time interval between two images also causes spatial misalignment. In order to simulate the misalignment between two images, we throw away the first few frames of optical flow map and only add up the following frames, as shown in Fig.~\ref{fig:blur_gen}.

After adding blur to the scaled irradiation image $sI_r$, we clip the range of the intensity to [0, 1]. Then, we sample the blurred irradiation image with a Bayer pattern to obtain a blurred raw image. The whole process to generate long-exposure raw images can be described as follows:
\begin{equation}
I_l = f_{clip}(f_{bayer}(f_{blur}(sI_r))+n),
\end{equation}
where $I_l$ is the synthetic long-exposed blur raw image; $f_{clip}$, $f_{bayer}$ and $f_{blur}$ are the clipping function, Bayer sampling function and blur function, respectively. $n$ is the noise, which is discussed later.

\subsection{Realistic noise}

Noise is inevitable in the imaging process, especially in low-light environment. In general, there are two main types of noise in raw images: shot noise and read noise. Shot noise occurs from individual photon detection events in the sensor, constituting a Possion process with variance equal to the signal level. Read noise is caused by a combination of sensor readout effects, and has an approximately Gaussian distribution. The overall noise can be described as a heteroscedastic Gaussian distribution:
\begin{equation}
n\sim \mathcal{N}(0, \sigma_s I + \sigma_r^2)
\end{equation}
where $I$ is the clean image. The noise parameters $\sigma_s$ and $\sigma_r$ are proportional to the ISO value. Since the exposure time of short exposure images is shorter than that of long-exposure images, the ISO values of short-exposure images need to be increased, or the image values need to be scaled, to match their brightness, which magnifies the noise variance of short-exposure images at the same time. In our method, the exposure ratio is 30, so the noise variance of short-exposure images is 30 times that of long-exposure images.

\begin{figure}[t]
\begin{center}
   \includegraphics[width=1\linewidth]{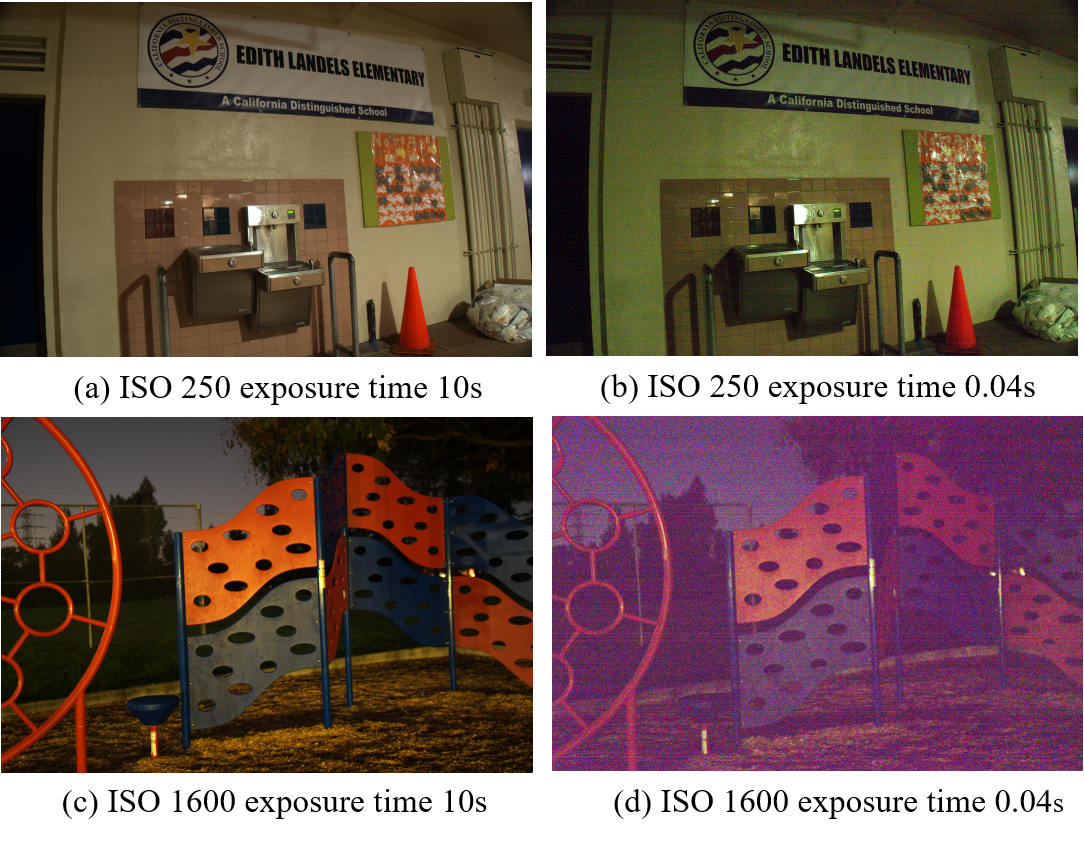}
\end{center}
   \caption{A typical example of color distortion. The low light images are chosen from SID dataset~\cite{chen2018learning}.}
\label{fig:color_distortion}
\end{figure}

\begin{figure*}[t]
\begin{center}
   \includegraphics[width=1\linewidth]{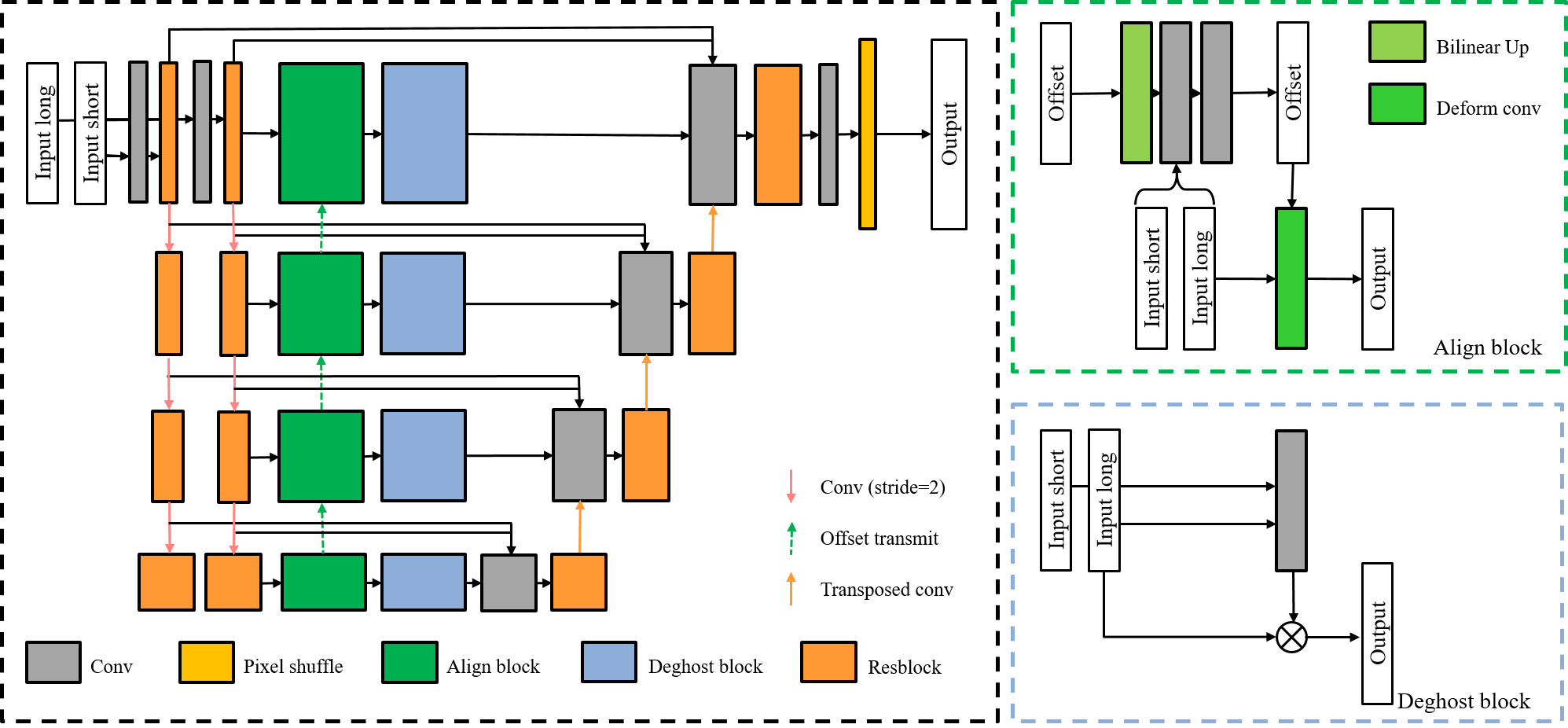}
\end{center}
   \caption{The architecture of our LSFNet.}
\label{fig:network}
\end{figure*}

\subsection{Color distortion}

The captured short-exposure image in a low-light environment often exhibits color distortion relative to the long-exposure images, especially when the exposure ratio is large. We assume that the ambient lighting is constant for each shot. In raw images recorded by sensors, green signals are higher than red and blue signals, so the estimated white balance coefficient is used to compensate for the red and blue signals in the ISP to restore the real scene color. Therefore, when the estimation of the white balance coefficient is inaccurate, the image color will be biased.

Another important cause of color distortion is the cutoff effect. Since the red and blue signals are smaller, underexposure cutoff is more likely to occur. When the noise variance of the image is large, the cutoff effect is more obvious. The cutoff effect causes the red and blue signal levels to be raised, thus shifting the underexposed images to purple.

A typical example of color distortion is shown in Fig.~\ref{fig:color_distortion}. We choose two pairs of short/long-exposure images from SID dataset~\cite{chen2018learning}, which are captured in a real low-light environment. When the noise level is relatively low, the short-exposure image is greenish. However, when the noise level is high, the short-exposure image tends to be purple due to the cutoff effect. Hence, to simulate this phenomenon, we multiply the red and blue channels $i$ of the irradiation image by a independent random coefficient $c_i$. Each $c_i$ is uniformly sampled from the range [0.7, 0.9]. After adding noise, we clip the image brightness and then perform white balancing on the clipped image.

\subsection{Synthesis of short-exposure raw images}
As mentioned above, we first reduce the brightness of the scaled irradiation image $sI_r$ by the exposure ratio $r$ to obtain the clean short-exposure images. Then, we introduce color distortion and noise. After Bayer sampling, we clip the brightness range to [0,1] to obtain the short raw images. The whole process to generate short-exposure raw images can be described as following:
\begin{equation}
I_s=f_{clip} (f_{bayer}(f_{color}(sI_r/r ))+n),
\end{equation}
where $f_{color}$ is the color distortion function. Before input into the network, the short-exposure image is scaled to match the brightness of the ground truth images.

\section{Model}

We design a novel long-short-exposure fusion network (LSFNet) to fuse the low-light raw pairs and fix images defects. The network takes the long- and short-exposure raw images as the input and makes use of their complementary advantages to improve image quality. The output is a three-channel image in the camera RGB domain. Then our network can be described as follows:
\begin{equation}
I_o={\rm LSFNet}(I_l, I_s),
\end{equation}
where $I_l$ and $I_s$ is long- and short-exposure raw images. $I_o$ is the output of the network, which is a camera RGB image. Therefore, we accomplish both fusion and demosaicing in the network. Then, we treat the output of the network into the ISP postprocessing to obtain a clear sRGB image. The same postprocessing operation is applied to the clipped and scaled irradiation image to generate the ground truth. We compute the loss in the sRGB domain to take into account the impact of ISP postprocessing.

\subsection{Network}

The architecture of our LSFNet is shown in Fig.~\ref{fig:network}. Our network structure consists mainly of four parts: feature extraction module, alignment module, deghosting module and fusion module. The whole network is designed as an encoder-decoder structure for multiscale fusion and image restoration. We will give more details on each modules in following.

\subsubsection{Feature extraction}
The feature extraction module extract useful features from the long and short exposure image respectively. The extracted features are used for image reconstruction after registration and enhancement. We first use a convolutional layer to extract features from the original scale. Then an encoder structure is used to extract multiscale information. In every scale, we use a Resblock~\cite{he2016deep} to extract the features of the current scale, and then use a convolution layer with a stride of 2 for downsampling.

\subsubsection{Alignment}
Since there is an offset between the short and long exposure inputs, alignment is needed before fusion and restoration. We assume that the short exposure image are spatially consistent to the ground truth image, so we mainly align the long exposure image to the short exposure image. The features of each location in the short exposure input needs to find the related features of the long exposure input for information complementation and fusion. To achieve this goal, we design the alignment blocks and applied them to the feature registration of each scale. We introduce deformable convolution~\cite{dai2017deformable, zhu2019deformable} into our alignment block. Different from the traditional convolution, deformable convolution can adaptively change the shape of the kernel to extract features related to the current location.  Deformable convolution has been applied to video quality restoration tasks~\cite{wang2019edvr,Chan2020UnderstandingDA} and its effectiveness has been proved. In the alignment block, we first use both short and long exposure inputs to calculate the offset maps and then reshape the features of the long exposure input. Furthermore, to better estimate the offset maps from coarse to fine, we transfer the last-scale offsets to the current scale to assist the offsets calculation. Therefore, the alignment block at the $t$ scale can be formulated as:
\begin{equation}
\Delta p^t = f_{cn}(f_{up}(\Delta p^{t-1}), f_{cn}(F_l^t, F_s^t)),
\end{equation}
\begin{equation}
F_l^t = f_{dcn}(F_l^t, \Delta p^t),
\end{equation}
where $F_l^t$ and $F_s^t$ is the long- and short-exposure features at the $t$ scale. $\Delta p$ is the offset map. $f_{cn}$ is the convolutional layer, and $f_{dcn}$ denotes the deformable convolution. $f_{up}$ denotes the upsampling function.

\subsubsection{Deghosting and feature enhancement}
After alignment, there may still be some areas where registration is inaccurate or misregistered, and some moving objects in the scene may cause ghost artifacts. In addition, Not every location feature has the same importance for image fusion. Therefore, we introduce the deghosting blocks to further suppress the ghost artifacts and enhance the features. The deghosting blocks are spatial attention mechanisms~\cite{yan2019attention, wang2019edvr}, which allow the network pay attention to the important features. They concatenate the two inputs and calculate a weight map, which lies in the range $[0,1]$,
\begin{equation}
W = f_{act}(f_{cn}(F_l, F_s)),
\end{equation}
where $f_{act}$ denotes the sigmoid activation function. $W$ denotes the weight map.
Then the weight map is used to reweight the long exposure features and suppresses the inconsistent areas,
\begin{equation}
F_l = F_l \circ W,
\end{equation}
We apply the deghosting blocks after the alignment blocks of each scale.

\subsubsection{Fusion and reconstruction}
We use an decoder structure to fuse the multiscale information and reconstruct the output. We first concatenate all the features and fuse them by a convolutional layer. Then the fused features are fed into a Resblock to reconstruct the output features. The features are reconstructed from coarse to fine using transposed convolutions upsampling.  At the end of the finest scale, we use a convolutional layer to recombine the features, and use a pixel-shuffle layer to reconstruct the three-channels output.

\subsubsection{Implement details}
We construct a four-scales network architecture with 32, 64, 128, 256 channels.  LeakyReLU~\cite{maas2013rectifier} with a slope of 0.2 is used as the activation function. The convolutional layers with a stride of 2 for downsampling and the transposed convolutional layers for upsampling use $2\times 2$ kernels. The last convolutional layer uses $1\times 1$ kernels. All other convolutional layers use $3\times 3$ kernels. The upsampling operation in offset transmission is bilinear interpolation.

\subsection{ISP postprocessing}
We perform white balancing and demosaicing before obtaining the output of the LSFNet. The output of the LSFNet lies in the camera RGB space, which is the same as the irradiation image. To obtain the visual sRGB image for display, we need ISP postprocessing including color space conversion and gamma correction. The color space conversion transforms the image into sRGB color space with a $3\times 3$ color correction matrix. Then, gamma correction improves the details in dark regions, making the image more consistent with human visual perception. We use the standard gamma curve:
\begin{equation}
\Gamma(I)=\max(I, \epsilon)^{1/2.22},
\end{equation}
where $\epsilon$, a constant to prevent numerical instability, is set to $10^{-8}$.

\subsection{HDR compression}
As mentioned above, the proposed network takes the short- and long-exposure raw pairs as input and outputs a sharp long-exposure image that is noise-free and has not motion blur. However, some overexposed areas still exist in the long-exposure image, thus limiting the dynamic range. Therefore, to maintain the details in highlighted areas, we retrain the network with dynamic range compression. The method of data generation and the architecture of the network are the same as described above. The only difference is the input of the network. We reduce the brightness of the long- and short-exposure images to match the brightness of the original irradiation image $I_r$ when input into the network. During ISP postprocessing, we replace the gamma correction with $\mu$-law\cite{kalantari2017deep} to compress the dynamic range, which is described as follows:
\begin{equation}
L=\frac{log(1+\mu H)}{log(1+\mu)},
\end{equation}
where $H$ is the HDR input image in the linear domain and $L$ is the tone-mapped output image. $\mu$ is a parameter which controls the ratio of compression. We set $\mu$ to 100 in our experiment. For the ground truth, we use the original irradiation image $I_r$ with the same ISP postprocessing, which is not scaled by $s$ and maintain normal exposure. In other words, we finish noise and motion blur removal, color correction, demosaicing and HDR compression in the proposed network, without other exposure fusion methods that would incur additional computational overhead.

\section{Experiments}
\subsection{Experimental setting}
We manually select 1322 high-quality raw images from the MIT-Adobe 5K dataset~\cite{fivek} to generate our training data. The selected images are sized approximately $2000\times 3000$ and have no obvious noise or motion blur. For the synthetic test data, we adopt 70 long-exposure raw images in the SID test dataset~\cite{chen2018learning}, which are around $3000\times 4000$ pixels. After that, we synthesize the training and test datasets using the method described in Section 4. For the real images, we capture pairs of short- and long-exposure raw images using the mobile phone. The long exposure time is set to 30 times the short exposure, and the ISO values for both are set the same.

We use the $L_1$ loss function for training the network, which is computed in the sRGB domain. The loss function can be described as:
\begin{equation}
L(\theta)=\frac{1}{N}\sum^N_{n=1}\|f_{isp}({\rm LSFNet}(I_l, I_s))-f_{isp}(sI_r)\|_1,
\end{equation}
where $\theta$ denotes the learned parameters in the network. $f_{isp}$ denotes the ISP postprocessing function.

In each training batch, we crop the raw images into $256\times 256$ patches and pack them into four R-G-G-B channels. Then, we use 16 patches with a size of $128\times 128\times 4\times 2$ as inputs. We train our model by ADAM optimizer~\cite{kingma2014adam} with $b_1=0.9$, $b_2=0.999$, and $e=10^{-8}$. The initial learning rate is set to $10^{-4}$ and then halved after 100 epochs. Our model is implemented by the PyTorch framework~\cite{paszke2019pytorch} and is trained using Nvidia GeForce RTX 1080Ti GPUs.

We compare several methods for image restoration, including denoising, deblurring and multi-image fusion methods. For the denoising methods, SID~\cite{chen2018learning} and RIDNet~\cite{anwar2019ridnet} are chosen, which are representative low-light denoising methods on raw and rgb domain respectively. The compared deblurring method is SRN~\cite{tao2018scale}. LSD$_2$~\cite{mustaniemi2018lsd} is chosen to represent the fusion method of long and short exposures.

\begin{figure}[t]
\begin{center}
   \includegraphics[width=1\linewidth]{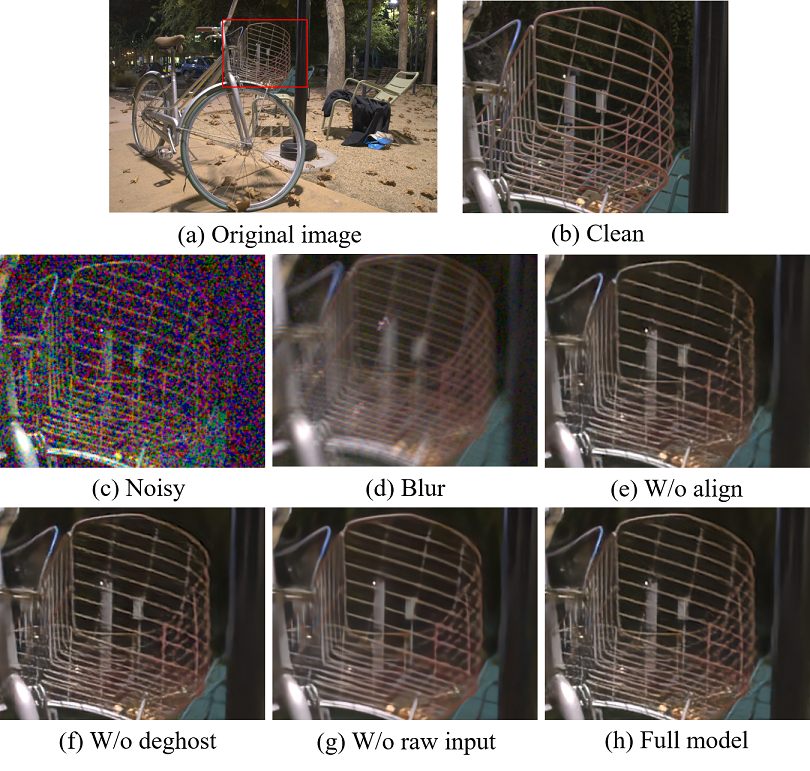}
\end{center}
   \caption{Ablation study of the proposed network on the synthetic dataset.}
\label{fig:ablation}
\end{figure}

\begin{figure*}[t]
\begin{center}
   \includegraphics[width=1\linewidth]{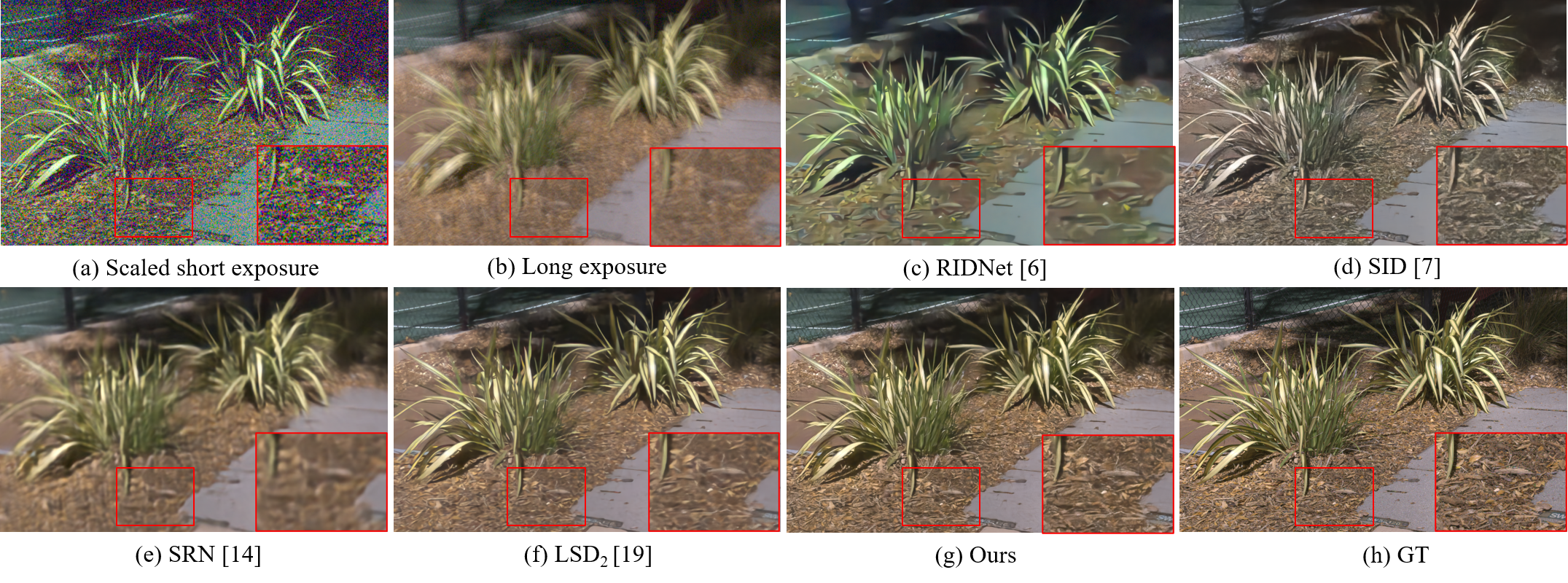}
\end{center}
   \caption{Results of the Grass image from the synthetic dataset.}
\label{fig:syn_res_1}
\end{figure*}

\begin{figure*}[t]
\begin{center}
   \includegraphics[width=1\linewidth]{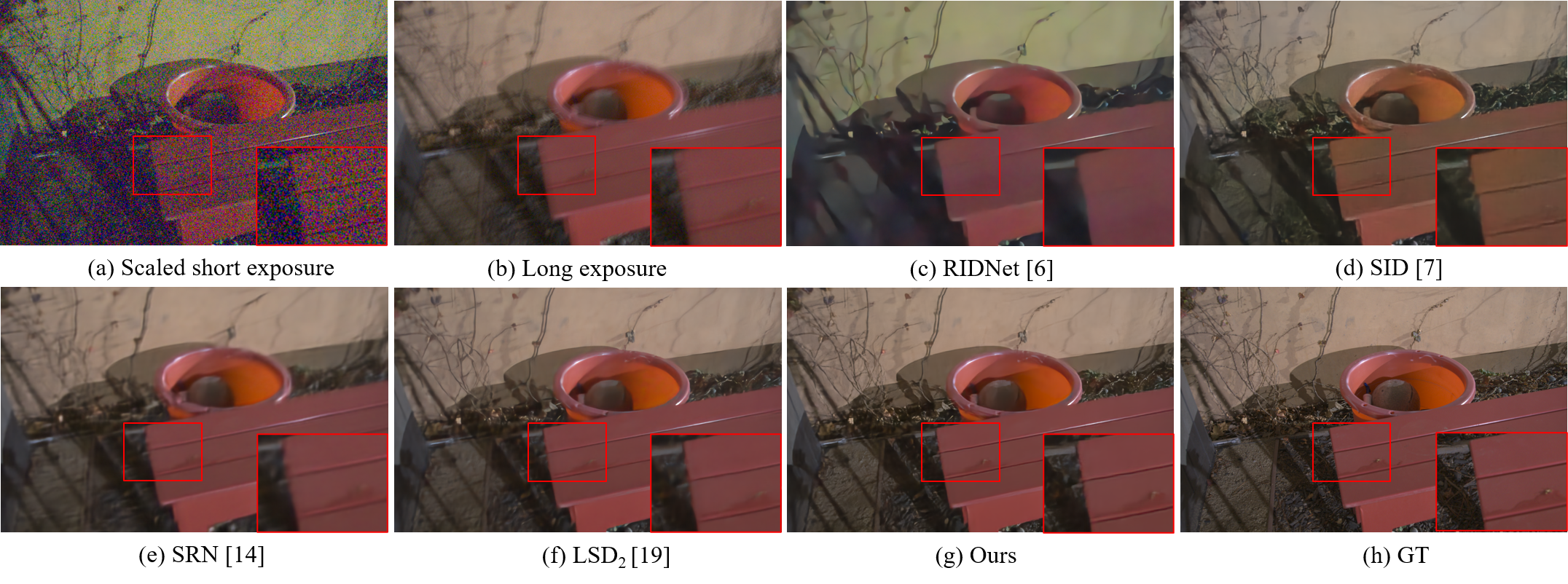}
\end{center}
   \caption{Results of the Chair image from the synthetic dataset.}
\label{fig:syn_res_2}
\end{figure*}

\subsection{Ablation study}
Since the proposed network contains multiple modules, we perform an ablation study on our synthetic test dataset to demonstrate their effectiveness. The quantitative results are provided in Table~\ref{table:ablation}, and the visual results are shown in Fig.~\ref{fig:ablation}. Without alignment modules, the network cannot effectively align the long- and short-exposure features, which results in many edges being smeared. Deghosting modules are complements to the alignment modules. The introduction of the deghosting modules further improves the details. Moreover, instead of taking raw images as the input directly, we input the sRGB images after ISP pipeline processing. The raw images are first white-balanced and demosaiced using the method proposed by~\cite{malvar2004high}. Then, the demosaiced images are processed by the ISP postprocessing as described above. The ISP pipeline makes noise distribution more complicated and compresses the image details. The restored images exhibit substantial loss of textures, and oversmoothing artifacts appear, which also result in a marked decline in quantitative indicators.

\begin{table}[!t]
\renewcommand{\arraystretch}{1.3}
\caption{Ablation study of different components.}
\label{table:ablation}
\begin{center}
\begin{tabular}{l|c|c|c|c}
\hline
Method & w/o align & w/o deghost & w/o raw input & full model \\
\hline\hline
PSNR & 30.05 & 30.56 & 30.10 & \textbf{30.70} \\
\hline
SSIM & 0.8735 & 0.8804 & 0.8725 & \textbf{0.8823}\\
\hline
\end{tabular}
\end{center}
\end{table}

\subsection{Comparison on synthetic images}

\begin{table}[!t]
\renewcommand{\arraystretch}{1.3}
\renewcommand\tabcolsep{25.0pt}
\caption{Quantitative comparison on the synthetic test dataset. The outputs of compared methods are adjusted to match the ground truth for fairness.}
\label{table:synthetic}
\begin{center}
\begin{tabular}{l|c|c}
\hline
Method & PSNR & SSIM \\
\hline\hline
RIDNet~~\cite{anwar2019ridnet} & 27.35 & 0.7846 \\
\hline
SRN~\cite{tao2018scale} & 26.24 & 0.7917 \\
\hline
SID~\cite{chen2018learning} & 27.53 & 0.7899\\
\hline
LSD$_2$~\cite{mustaniemi2018lsd} & 29.97 & 0.8480\\
\hline
LSFNet (ours) & \textbf{30.70} & \textbf{0.8602}\\
\hline
\end{tabular}
\end{center}
\end{table}

For a fair comparison on synthetic images, we retrain all the methods on our synthetic training data.
The quantitative results for the synthetic images are listed in Table~\ref{table:synthetic}. Before the comparison, we align the deblurring results of SRN~\cite{tao2018scale} with the ground truth. We also adjust the colors of the RIDNet~\cite{anwar2019ridnet} outputs to match the ground truth. In the quantitative comparison, both PSNR and SSIM values of our method exceed those of all the other methods in the comparison.

Fig.~\ref{fig:syn_res_1} and Fig.~\ref{fig:syn_res_2} show some examples from our synthetic dataset. We extend the brightness of the short-exposure images for better display. RIDNet~\cite{anwar2019ridnet} can handle real noise on sRGB images. However, when the noise of short-exposure images is high, many textures are smeared and high-frequency information is lost.  In addition, RIDNet~\cite{anwar2019ridnet} does not consider the color distortion when denoising and cannot correct the color of short-exposure images. SID~\cite{chen2018learning} removes noise from raw images and considers the whole ISP pipeline in an end-to-end network. However, limited by the information from short-exposure images, SID~\cite{chen2018learning} cannot restore the rich details. Without long-exposure images as references, SID~\cite{chen2018learning} often makes mistakes in color correction and fails to generalize to different imaging devices.

SRN~\cite{tao2018scale} can deal with motion blur when the scene is relatively simple. However, many artifacts appear in the results. The resulting scenes are still over-smoothing, and many details are lost, such as the leaves in Fig.~\ref{fig:syn_res_1}, or lead to ghosts as shown in Fig.~\ref{fig:syn_res_2}. Motion blur in some areas cannot be removed and many textures are smoothed. The existence of noise also worsens the deblurring results and leads to artifacts. Similar to our method, LSD$_2$~\cite{mustaniemi2018lsd} utilizes the complementary information of long and short exposures, which can remove noise and blur simultaneously and restore true color. However, limited by the network and ISP postprocessing, some details are smoothed. In contrast, our method can restore vivid textures and sharp edges without other artifacts.

\begin{figure}[t]
\begin{center}
   \includegraphics[width=1\linewidth]{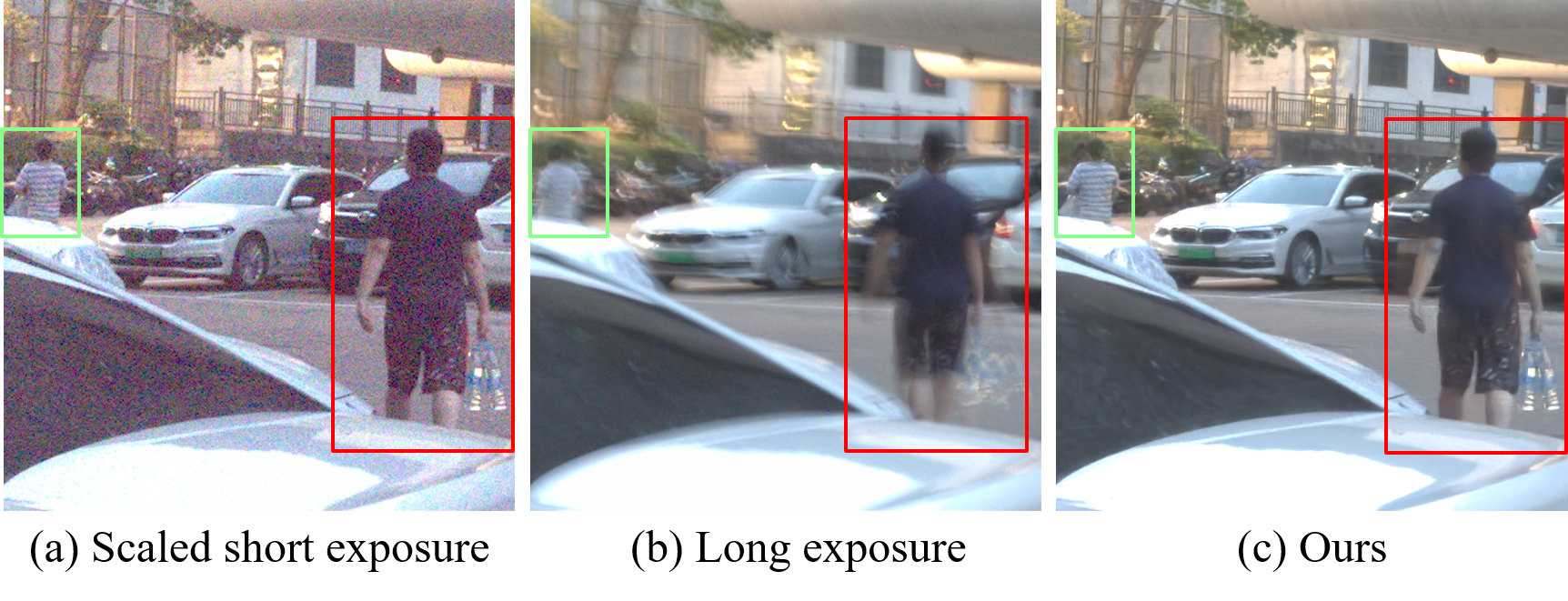}
\end{center}
   \caption{Object movement on real scene.}
\label{fig:object}
\end{figure}

\begin{figure*}[t]
\begin{center}
   \includegraphics[width=1\linewidth]{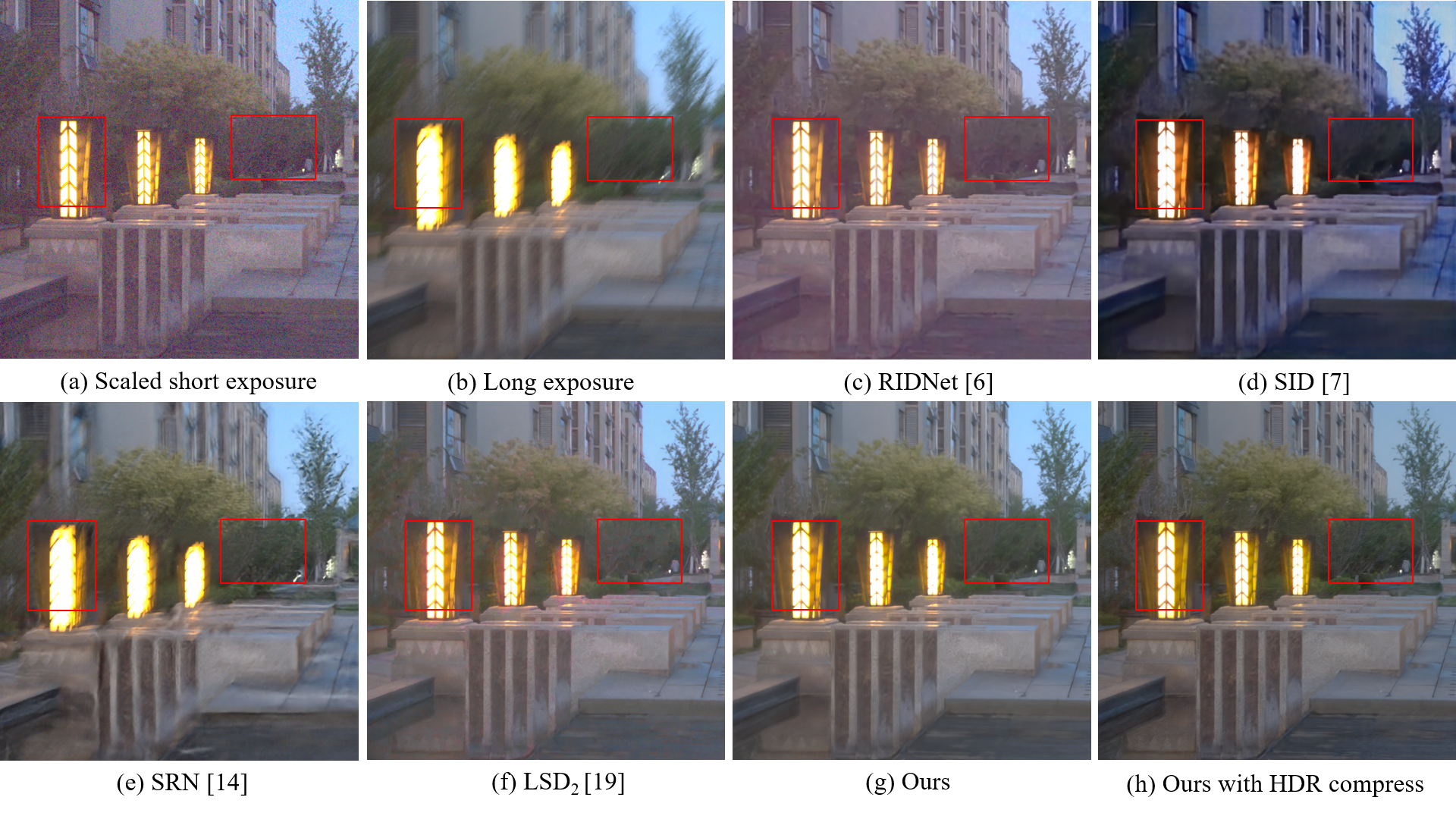}
\end{center}
   \caption{Results of the Lamp image in a real scene.}
\label{fig:real_res_1}
\end{figure*}

\begin{figure*}[t]
\begin{center}
   \includegraphics[width=1\linewidth]{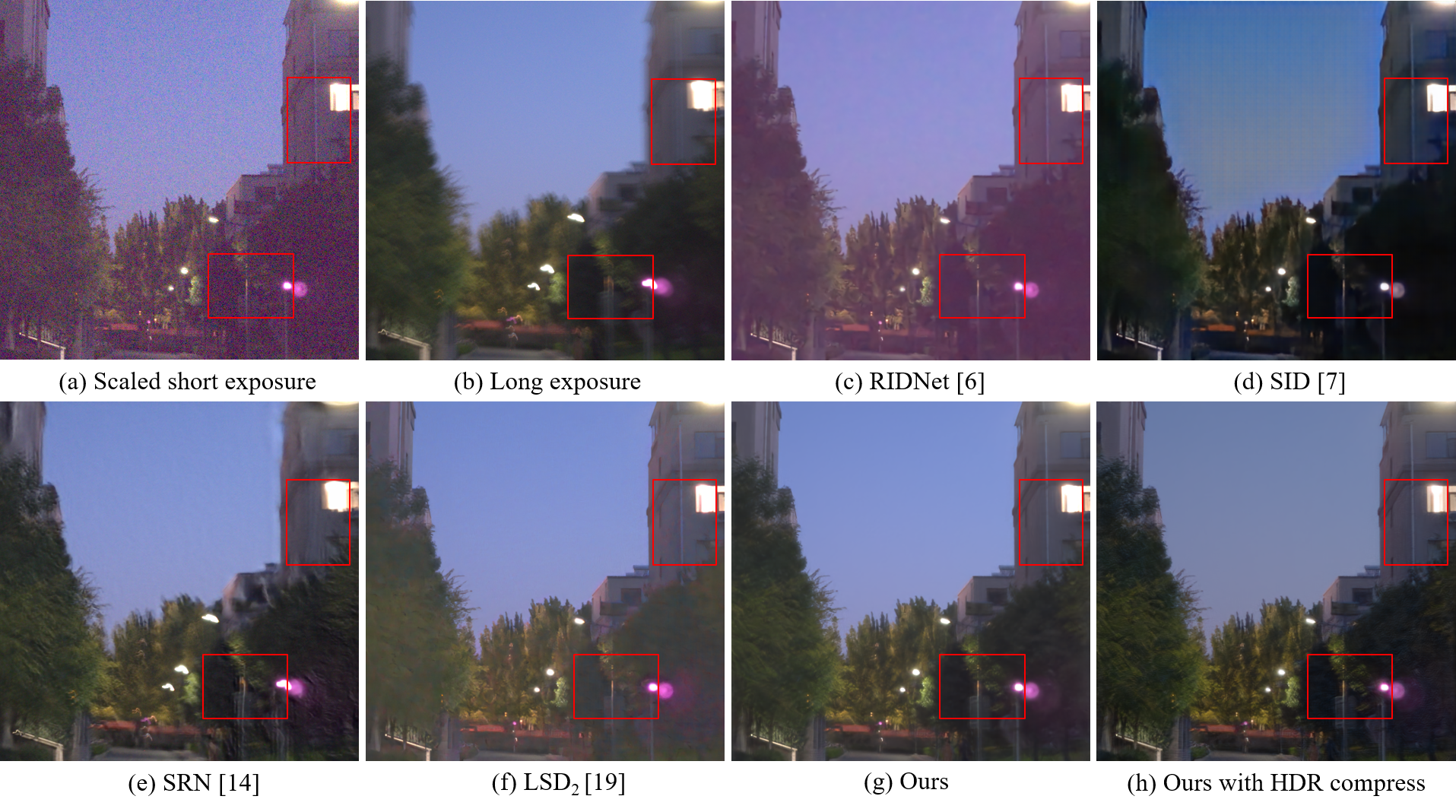}
\end{center}
   \caption{Results of the Street image in a real scene.}
\label{fig:real_res_2}
\end{figure*}

\subsection{Comparison on real images}

We collected some real world images using an oneplus 7T mobile phone. we modify the burst mode in the camera of the mobile phone so that it could consecutively capture a pair of short- and long-exposure raw images. The long exposure time is set to 1 second and The short exposure time is set 1/30 second. The ISO values for both raw images are set 100. The captured raw images are saved as DNG format.

We capture some real low-light scenes and evaluate our method on these real images. Our method uses the edges of the short-exposure image as a constraint so that the resulting scene is consistent with the short exposure image. Therefore, our method can get rid of motion blur caused by both camera shake and object movement, as long as the short exposure image is blur-free. As shown in Figs.~\ref{fig:object}, in addition to the blur caused by camera shake, there are also movement of objects in the scene, such as the pedestrians in red and green boxes. Our method can obtain clean result without motion blur and noise.

Figs.~\ref{fig:real_res_1} and \ref{fig:real_res_2} show other examples on real scenes. Without long exposure as a  reference, the denoising methods cannot recover the color correctly. The overexposed areas are enlarged in the blurred images and the deblurring methods such as SRN~\cite{tao2018scale} cannot restore the cutoff details. LSD$_2$~\cite{mustaniemi2018lsd} cannot remove the noise in textured areas such as grass and leaves. Moreover, ghosts may appear in highlighted areas due to misalignment, as shown in Fig.~\ref{fig:real_res_1}. In contrast to other method, our method obtains rich textures. We also show the results of our method with HDR compression. The overexposed areas in the long exposure images are restored, and high contrast is maintained in other areas.

\begin{figure}[t]
\begin{center}
   \includegraphics[width=1\linewidth]{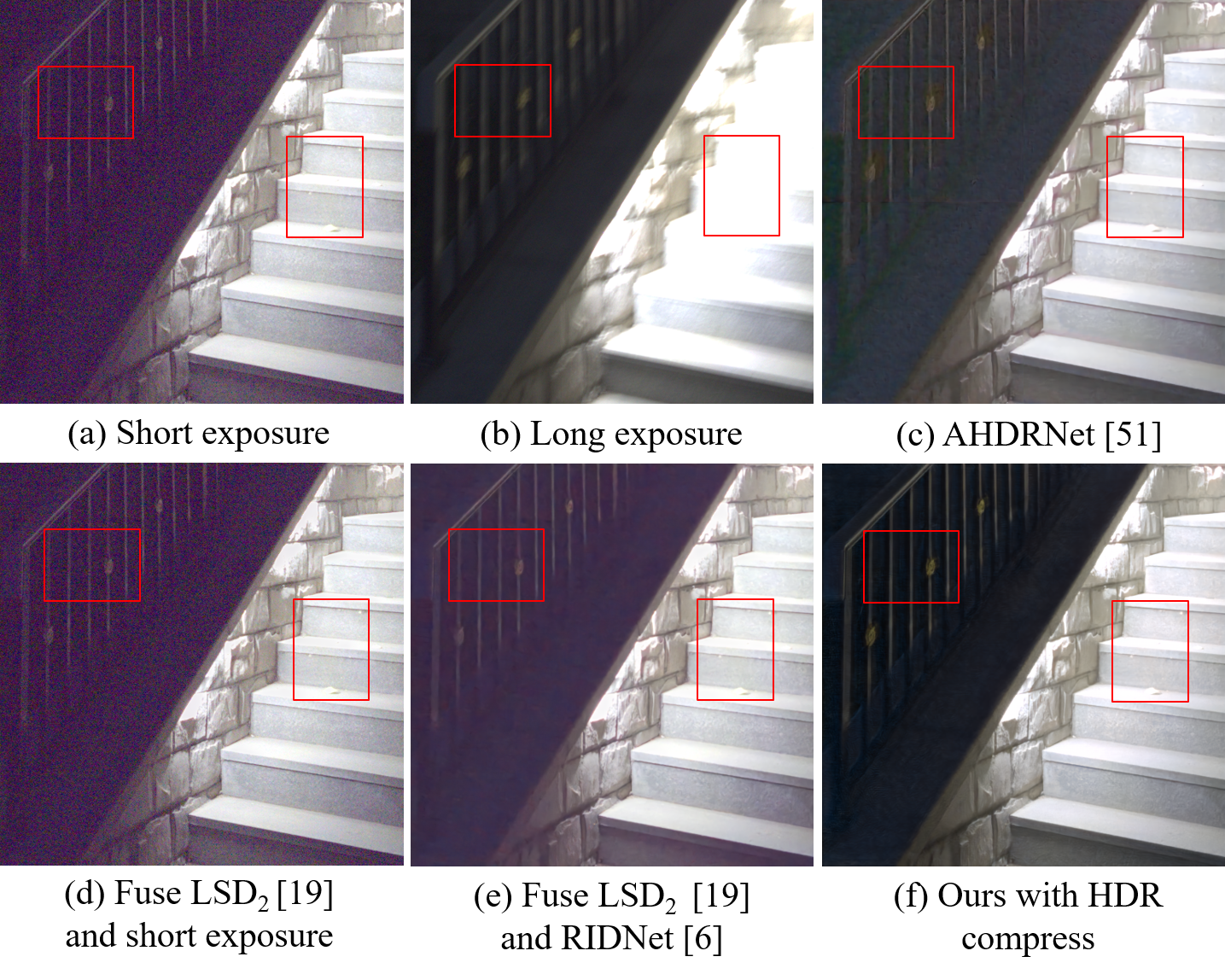}
\end{center}
   \caption{Comparison with exposure fusion on real scene.}
\label{fig:fusion}
\end{figure}

\subsection{Comparison with exposure fusion}
Other exposure fusion methods improve the dynamic range by fusing multi-exposure images only. They assume that the input images are of high quality with only the dynamic range cut off. When the input images have noise or other artifacts, they may magnify them with the textures in the outputs. As shown in Fig.~\ref{fig:fusion}, we fuse the long-exposure image of LSD$_2$'s output~\cite{mustaniemi2018lsd} and short-exposure image using exposure fusion~\cite{Mertens2007Exposure}. The fusion result has a high noise level. Although the short-exposure image is denoising using RIDNet~\cite{anwar2019ridnet}, some textures are lost. The fusion result still has low contrast and poor details. In addition, we also test AHDRNet~\cite{yan2019attention} as a representative of the CNN-based exposure fusion method. To make a fair comparison, we retrain AHDRNet on our training data with same u-law HDR compression. However, it leads to noise residual and artifacts in low light areas. In contrast, our LSFNet can output a sharp tone-mapped result directly without additional exposure fusion.

\begin{figure}[t]
\begin{center}
   \includegraphics[width=1\linewidth]{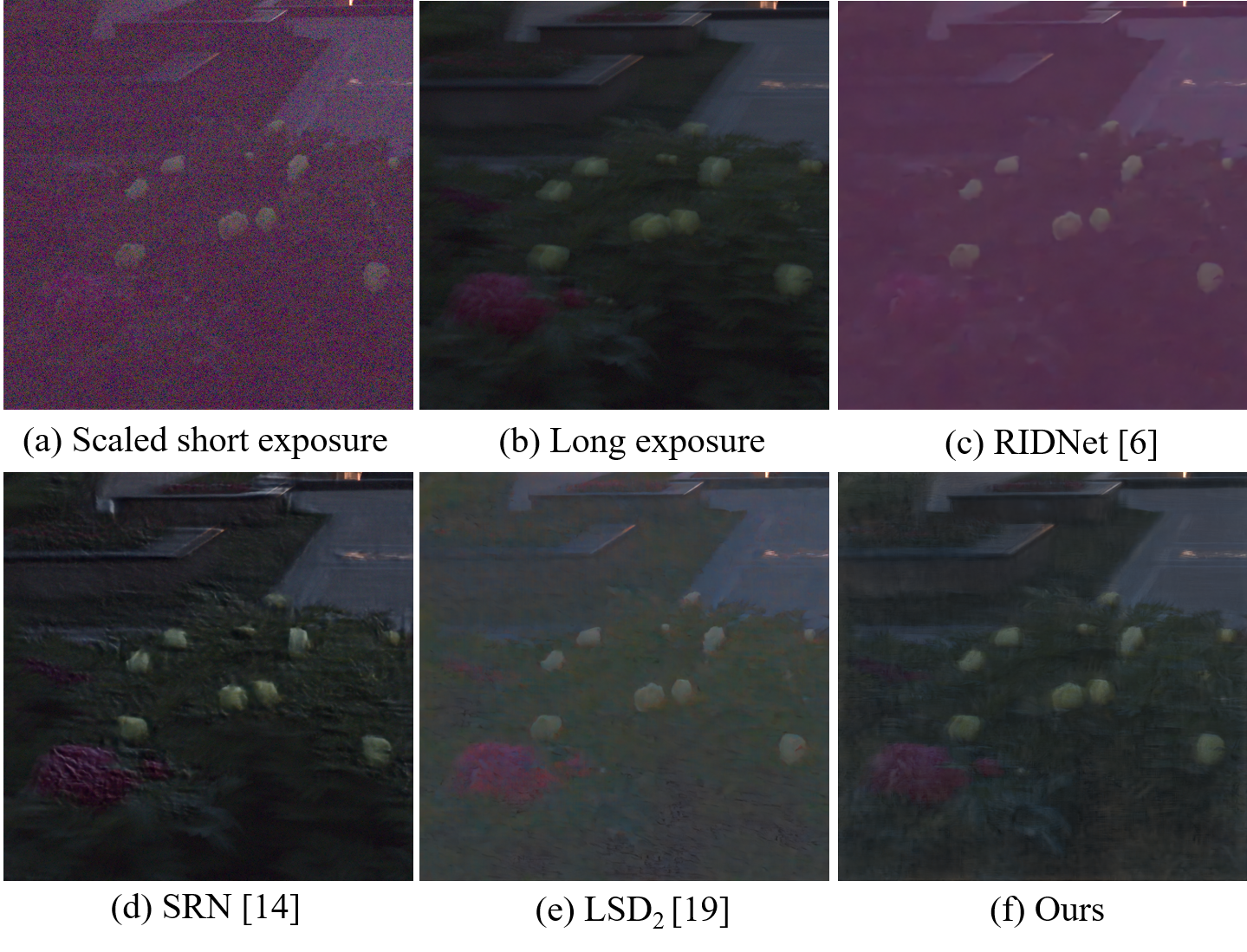}
\end{center}
   \caption{An extreme imaging case in a real scene.}
\label{fig:limit}
\end{figure}

\subsection{Computational overhead}

\begin{table}[!t]
\renewcommand{\arraystretch}{1.3}
\renewcommand\tabcolsep{9pt}
\caption{Parameter number and time comparisons on 256$\times$256 input patches.}
\label{table:para}
\begin{center}
\begin{tabular}{l|c|c|c|c}
\hline
Method  & Params & FLOPs & Times (s) & PSNR (dB)  \\
\hline\hline
SRN~\cite{tao2018scale} & 10.3M & 108.6G & 0.062 & 26.23\\
\hline
SID~\cite{chen2018learning} &  7.8M &  13.8G & 0.005 & 27.53\\
\hline
RIDNet~\cite{anwar2019ridnet} &  1.5M &  98.1G & 0.039 & 27.34\\
\hline
LSD$_2$~\cite{mustaniemi2018lsd} &  31.0M & 54.8G &  0.011 &  29.97\\
\hline
LSFNet (ours) &  8.4M & 38.0G & 0.019 &  30.70 \\
\hline
\end{tabular}
\end{center}
\end{table}

We test different methods on 256$\times$256 input patches to compare the computational overhead. All the methods are implemented in PyTorch. We provide the number of floating point operations (FLOPs) since the running time may depend on the test platform and code. As shown in Table~\ref{table:para}, our method achieves the optimal performance with a moderate computational overhead.

\section{Discussion and Limitations}
Although our method can produce high-quality images in real scenes, it has several limitations that may inspire future work.

Our method adopts the complementary information of short- and long-exposure images. However, the complementary information is missing when the quality of captured images is poor. On the one hand, short-exposure images are too noisy to distinguish details; useful information come from long exposure images only, and the model degenerates into a single-image deblurring network in this case. On the other hand, when long-exposure images are extremely blurred, the quality of the outputs also deteriorates. However, as long as the noise of short-exposure images is not severe, the correct color and textures can still be recovered. In some harsh imaging environments, such as extremely low-light condition, the captured image pairs have high noise and large motion blur, which leads to the difficulty in outputting satisfactory results.

As shown in Fig.~\ref{fig:limit}, both a high noise level and a large motion blur exist in the input image pair. The proposed method fails to recover sharp result, especially in the textured areas where many artifacts appear. However, under such conditions, other methods also have difficulty achieving good results. More input images or other support information may be needed to improve the image quality.

\section{Conclusion}
In this paper, we propose a novel low-light restoration method that fuses a noisy short-exposure image and a blurred long-exposure image into a high quality sharp image. Based on simulating the imaging process in a low-light environment, a new data generation method is proposed to synthesize more realistic raw images with a variety of exposures. Moreover, we design a new network to handle the short- and long-exposure raw images and output an RGB image without noise, blur or color distortion. We compare various low-light image enhancement methods and demonstrate that our method can obtain the state-of-the-art performance.

\section*{Acknowledgments}

This work was supported by Basic Research on Civil Aerospace No.D040301.

\ifCLASSOPTIONcaptionsoff
  \newpage
\fi


\bibliographystyle{IEEEtran}
\bibliography{reference}




\end{document}